\documentclass{emulateapj}
\usepackage{times}

\newcommand{\gray}{$\gamma$-ray}
\newcommand{\grays}{$\gamma$-rays}
\newcommand{\sigv}{\langle\sigma_a v\rangle}
\newcommand{\pubjournal}[6] {#1 #5, #2, {\bf #3}, #4}
\newcommand{\pubbook}[6]{#1, #5, #2 #3, #4}

\shorttitle{Dark matter burners: Preliminary estimates}
\shortauthors{Moskalenko \& Wai}

\begin{document}
\frenchspacing

\title{Dark matter burners: Preliminary estimates}
\author{Igor V. Moskalenko\altaffilmark{1}}
\affil{
   Hansen Experimental Physics Laboratory,
   Stanford University, Stanford, CA 94305 \email{imos@stanford.edu}}
\altaffiltext{1}{Also Kavli Institute for Particle Astrophysics and Cosmology,
Stanford University, Stanford, CA 94309}
\and
\author{Lawrence Wai\altaffilmark{1}}
\affil{Stanford Linear Accelerator Center, Stanford University, 
2575 Sand Hill Rd, Menlo Park, CA 94025 \email{wai@slac.stanford.edu}}

\begin{abstract}

We show that a star orbiting close enough to an adiabatically grown
supermassive black hole can capture a large number of weakly
interacting massive particles (WIMPs) during its lifetime.  WIMP
annihilation energy release in low- to medium-mass stars is comparable
with or even exceeds the luminosity of such stars due to thermonuclear
burning.  The excessive energy release in the stellar core may result
in an evolution scenario different from what is expected for a regular
star.  The model thus predicts the existence of unusual stars within
the central parsec of galactic nuclei.  If found, such stars would
provide evidence for the existence of particle dark matter.
White dwarfs seem to be the most promising candidates to look for.
The signature of a white dwarf burning WIMPs
would be a very hot star with mass and radius characteristic for a white 
dwarf, but with luminosity exceeding the typical luminosity of a white 
dwarf by orders of magnitude $\la$$50L_\odot$. A white dwarf with a
highly eccentric orbit around the central black hole may exhibit
variations in brightness correlated with the orbital phase.  

\end{abstract}


\section{Introduction}

The nature of the non-baryonic dark matter, which dominates the
visible matter by about 4:1, is perhaps the most interesting
experimental challenge for contemporary particle astrophysics.  A hint
for a solution has been found in particle physics where the WIMPs
arise naturally in supersymmetric extensions of the Standard Model
\citep[e.g.,][]{haber-kane}, among other possibilities.  The WIMP is
typically defined as a stable, electrically neutral, massive particle.
Assuming that non-baryonic dark matter is dominated by WIMPs, the pair
annihilation cross-section is restricted to provide the observed relic
density \citep{jkg96,bergstrom00}.  A pair of WIMPs can annihilate
producing ordinary particles and \grays.

WIMPs are expected to form high density clumps according to N-body
simulations of test particles with only gravitational interactions
\citep{Navarro97,Moore99}.  The highest density ``free space'' dark
matter regions occur for dark matter particles captured within the
gravitational potential of adiabatically grown supermassive black
holes \citep{gs99,bm05}.  Higher dark matter densities are possible for
dark matter particles captured inside of stars or planets.  Any star
close enough to an adiabatically grown supermassive black hole can
capture a large number of WIMPs during a short period of time.

Such an idea has been first proposed by \citet{salati89} and
further developed by \citet{bouquet89} who applied it to main-sequence
stars.  WIMP annihilation in stars may lead to considerable energy
release in the stellar cores thus affecting the evolution and
appearance of such stars.  The model led to the conclusion of
suppression of stellar core convection, thus predicting a
concentration of stars in the Galactic center masquerading as cold red
giants.

We perform order-of-magnitude estimates of WIMP capture rates for
stars of various masses and evolution stages located in high
density dark matter regions. We use current limits on WIMP-nucleus
interaction and WIMP annihilation cross sections, as well as recent
estimates of WIMP energy density near an adiabatically grown
supermassive black hole. We argue that white dwarfs, fully burned
stars without their own energy supply, are the most promising
candidates to look for.

\section{WIMP accumulation in stars}

In a steady state the WIMP capture rate $C$ is balanced by the
annihilation rate \citep{gs87}
\begin{equation}
C=A N_\chi^2
\label{balance}
\end{equation}
where $N_\chi$ is the total number of WIMPs in the star,
\begin{eqnarray}
&&A=\sigv/V_{\rm eff},
\label{A}\\
&&V_{\rm eff}=\pi^{3/2} r_\chi^3,
\label{Veff}
\end{eqnarray}
$\sigv$ is the velocity averaged WIMP pair annihilation
cross-section, and the effective volume $V_{\rm eff}$ is 
determined by matching the core temperature $T_c$ with the 
gravitational potential energy at the core radius $r_\chi$
(assuming thermal equilibrium). The total number of captured WIMPs is
\begin{equation}
N_\chi=C \tau_{eq} \tanh(\tau_*/\tau_{eq}),
\label{Nchi}
\end{equation}
where $\tau_*$ is the star age, and the equilibrium time scale is given by
\begin{equation}
\tau_{eq}=(CA)^{-1/2}.
\label{taueq}
\end{equation}
The number density distribution of WIMPs can be estimated as 
\citep{ps85,gs87,bottino02}:
\begin{equation}
n_\chi(r)=n_\chi^c \exp(-r^2/r_\chi^2),
\label{chidensity}
\end{equation}
where $n_\chi^c=N_\chi/V_{\rm eff}$ is the
central WIMP number density. In thermal equilibrium, 
the effective radius $r_\chi=r_T$ is determined by the 
core temperature $T_c$ and density $\rho_c$
\begin{equation}
r_T=c \left(\frac{3T_c}{2\pi G\rho_c m_\chi}\right)^{1/2},\nonumber
\label{rT}
\end{equation}
where $c$ is the speed of light,
$G$ is the gravitational constant, and $m_\chi$ is the WIMP mass. 

\begin{figure}[t]
\centerline{
\includegraphics[width=0.48\textwidth]{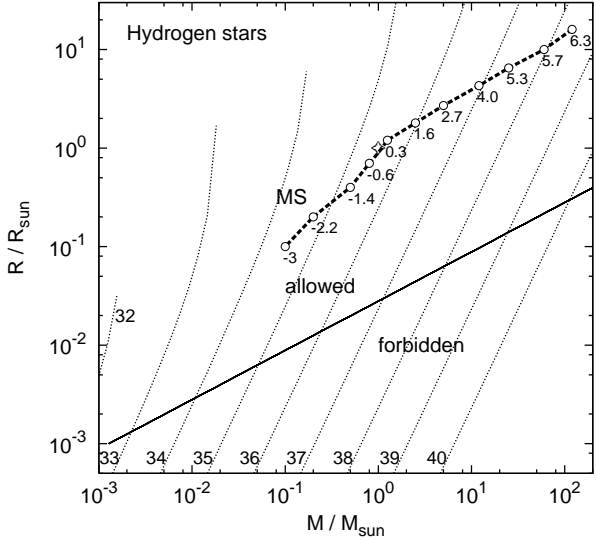}}
\caption{Mass--radius--capture rate diagram for Hydrogen stars.  The
dotted line series are calculated using eq.~(\ref{C2}) and the numbers
on the plot show the logarithm of WIMP capture rate $\log_{10} C$.
The strong dashed line labeled ``MS'' shows characteristics of the
main sequence stars \citep{encyclopedia}, and the diamond symbol
indicates a sun-like star with $M=M_\odot$, $R=R_\odot$. The open
circles on the main sequence show the stars of particular spectral
type and the numbers show their luminosity $\log_{10} (L_*/L_\odot)$.
The solid line divides the diagram into ``allowed'' and ``forbidden''
parts, see text for more details.
\label{mapH}}
\vspace{-1em}
\end{figure}

If the WIMP-nucleon scattering cross-section is large enough, then
every WIMP crossing the star is captured and $C$ saturates to a
maximal value proportional to the star's cross-sectional area $\pi
R_*^2$, the WIMP velocity dispersion $\bar{v}$, and the ambient WIMP
energy density $\rho_\chi$:
\begin{equation}
C = \left(\frac{8}{3\pi}\right)^{1/2} \frac{\rho_\chi \bar{v}}{m_\chi}
\left[ \zeta+\frac{3 v_{esc}^2}{2\bar{v}^2} \right] \pi R_*^2,
\label{C1}
\end{equation}
where $\zeta$ is taken to be 1.77, $v_{esc}=\sqrt{2GM_*/R_*}$ is the
escape velocity at the star's surface, and $M_*$ is the star mass.

On the other hand, limits from direct detection of dark matter on the
WIMP-nucleon cross-section imply that only a fraction of the WIMPs
crossing the star will scatter and be captured.  In this case the
star's cross-sectional area is replaced by the product of the
WIMP-nucleon scattering cross-section $\sigma_n$, the number of
nuclei, and a spin-independent coherent factor for nuclei heavier than Hydrogen, that
is $A_n^4$ for a nucleus with the atomic mass $A_n$. The limits on
spin-independent WIMP-nucleon scattering are more stringent than for
spin-dependent; from direct detection experiments the spin-independent
WIMP-nucleon scattering cross-section $\sigma_0$ is at most $10^{-43}$
cm$^2$ \citep{cdms06}.  The current limit on spin-dependent Hydrogen
scattering comes from Super-Kamiokande, which finds that $\sigma_s$ is
at most $10^{-38}$ cm$^2$ \citep{sk04}. In the following formulas we
will use a generic cross-section $\sigma_n$ to be substituted by
$\sigma_0$ or $\sigma_s$ for spin-independent and spin-dependent
interactions, correspondingly.  If the star is entirely composed of
nuclei with the atomic mass $A_n$ ($A_n=1$ for pure Hydrogen), we have
for $m_\chi\gg A_n m_p$:
\begin{equation}
C = \left(\frac{8}{3\pi}\right)^{1/2} \frac{\rho_\chi \bar{v}}{m_\chi}
\left[ \zeta+\frac{3 v_{esc}^2}{2\bar{v}^2} \right] 
\sigma_n A_n^4 \frac{M_*}{A_n m_p},
\label{C2}
\end{equation}
where $m_p$ is the proton mass.

Eq.~(\ref{C1}) gives an absolute upper limit on the capture rate.
The allowable range of masses and radii at a given WIMP-nucleon 
scattering cross section can be obtaned from eqs.~(\ref{C1})
and (\ref{C2}):
\begin{equation}
R_*\ge A_n^{3/2}\sqrt{\frac{\sigma_n M_*}{\pi m_p}}.
\label{MR}
\end{equation}
Smaller radii are forbidden as the capture rate will
exceed the upper limit given by eq.~(\ref{C1}). 

\begin{figure}[t]
\centerline{
\includegraphics[width=0.48\textwidth]{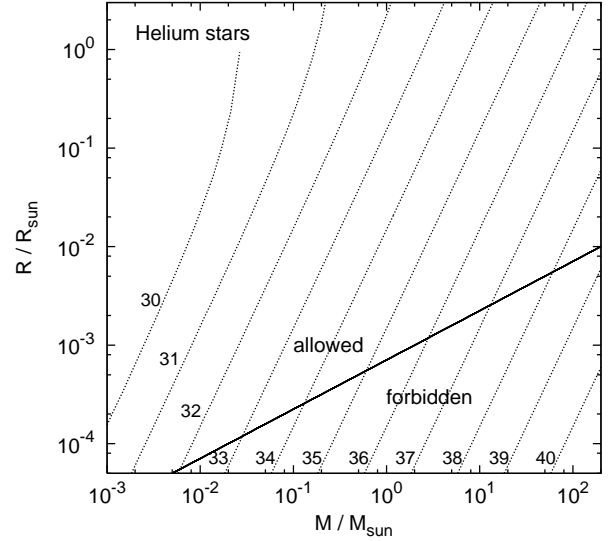}}
\caption{Mass--radius--capture rate diagram for Helium stars.
Lines are coded as in Figure~\ref{mapH}.
\label{mapHe}}
\vspace{-1em}
\end{figure}


\begin{deluxetable*}{ccccccccccc}
\tablecolumns{11}
\tablewidth{0pt}
\tabletypesize{\scriptsize}
\tablecaption{WIMP accumulation during different burning stages
\label{table1}}
\tablehead{
\colhead{$M^H_{\rm init}$} &
\colhead{$M_*$} &
\colhead{$R_*$} &
\colhead{$T_c$} &
\colhead{$\rho_c$} &
\colhead{$C$} &
\colhead{$r_{th}$} &
\colhead{$\tau_{eq}$} &
\colhead{$\tau_*$} &
\colhead{$N_\chi$} &
\colhead{$n_\chi^c$}
\smallskip\\
\colhead{$M_\sun$} &
\colhead{$M_\sun$} &
\colhead{$R_\sun$} &
\colhead{$10^8$ K} &
\colhead{g cm$^{-3}$} &
\colhead{s$^{-1}$} &
\colhead{$10^8$ cm} &
\colhead{s} &
\colhead{s} & 
\colhead{} &
\colhead{cm$^{-3}$} 
}
\startdata
\multicolumn{2}{c}{Hydrogen Stage}
\smallskip\\

1                   & 
1                   & 
1                   & 
0.157               & 
153                 & 
$2.85\times10^{35}$ & 
7.55                & 
$5.29\times10^{ 8}$ & 
$3.47\times10^{16}$ & 
$1.51\times10^{44}$ & 
$6.30\times10^{16}$   
\smallskip\\

13                  & 
12.9                & 
6.24                & 
0.344               & 
6.66                & 
$6.88\times10^{36}$ & 
53.5                & 
$2.04\times10^{ 9}$ & 
$4.26\times10^{14}$ & 
$1.40\times10^{46}$ & 
$1.64\times10^{16}$   
\smallskip\\

25                  & 
24.5                & 
9.17                & 
0.381               & 
3.81                & 
$1.65\times10^{37}$ & 
74.5                & 
$2.16\times10^{ 9}$ & 
$2.11\times10^{14}$ & 
$3.56\times10^{46}$ & 
$1.55\times10^{16}$   
\smallskip\\

75                  & 
67.3                & 
21.3                & 
0.426               & 
1.99                & 
$5.30\times10^{37}$ & 
109                 & 
$2.13\times10^{ 9}$ & 
$9.97\times10^{13}$ & 
$1.13\times10^{47}$ & 
$1.56\times10^{16}$   
\smallskip\\

75\tablenotemark{a} & 
75                  & 
9.36                & 
0.760               & 
10.6                & 
$1.44\times10^{38}$ & 
63.1                & 
$5.69\times10^{ 8}$ & 
$1.09\times10^{14}$ & 
$8.19\times10^{46}$ & 
$5.85\times10^{16}$   
\medskip\\

\multicolumn{2}{c}{Helium Stage}
\smallskip\\

 1 &  0.71 & 10  & 1.25 & 20000 & $3.13\times10^{31}$ & 1.86 & $6.19\times10^{9}$ & $3.47\times10^{15}$ & $1.94\times10^{41}$ & $5.39\times10^{15}$
\smallskip\\
13 & 12.4 & 359  & 1.72 & 1730  & $4.80\times10^{32}$ & 7.43 & $1.26\times10^{10}$ & $8.43\times10^{13}$ & $6.05\times10^{42}$ & $2.65\times10^{15}$
\smallskip\\
25 & 19.6 & 1030 & 1.96 & 762   & $7.14\times10^{32}$ & 11.9 & $2.11\times10^{10}$ & $2.65\times10^{13}$ & $1.50\times10^{43}$ & $1.58\times10^{15}$
\smallskip\\
75 & 16.1 & 1.17 & 2.10 & 490   & $3.35\times10^{34}$ & 15.4 & $4.51\times10^{9}$ & $1.51\times10^{13}$ & $1.51\times10^{44}$ & $7.40\times10^{15}$
\smallskip\\
75\tablenotemark{a} & 
     74.4 & 702  & 2.25 & 319   & $5.74\times10^{36}$ & 19.8 & $5.01\times10^{ 8}$ & $1.05\times10^{13}$ & $2.87\times10^{45}$ & $6.66\times10^{16}$
\medskip\\

\multicolumn{2}{c}{Carbon Stage}
\smallskip\\

$\sim$3&  1.0 & 0.01 & 10   & $1.00\times10^{6}$ & $4.03\times10^{35}$ & 0.745 & $1.38\times10^{7}$ &   \nodata  & $5.56\times10^{42}$ & $2.41\times10^{18}$
\smallskip\\
13     & 11.4 &  665 & 8.15 & $3.13\times10^{5}$ & $1.11\times10^{34}$ & 1.20 & $1.70\times10^{8}$ & $8.90\times10^{10}$ & $1.89\times10^{42}$ & $1.96\times10^{17}$
\smallskip\\
25     & 12.5 & 1390 & 8.41 & $1.29\times10^{5}$ & $1.18\times10^{34}$ & 1.90 & $3.29\times10^{8}$ & $1.65\times10^{10}$ & $3.88\times10^{42}$ & $1.01\times10^{17}$
\smallskip\\
75     & 6.37 & 0.644& 8.68 & $1.39\times10^{5}$ & $2.59\times10^{35}$ & 1.86 & $6.80\times10^{7}$ & $3.38\times10^{10}$ & $1.76\times10^{43}$ & $4.90\times10^{17}$
\smallskip\\
75\tablenotemark{a} &
         74.0 & 714  & 10.4 & $7.45\times10^{4}$ & $5.67\times10^{36}$ & 2.78 & $2.66\times10^{7}$ & $8.52\times10^{8}$ & $1.51\times10^{44}$ & $1.25\times10^{18}$
\smallskip

\enddata
\tablenotetext{a}{Stellar model with $10^{-4}$ solar metallicity and assuming a massive Hydrogen envelope ($\sigma_n=\sigma_s$).}
\tablecomments{$M^H_{\rm init}$ is the initial hydrogen star mass.}
\end{deluxetable*}

A star spends about 90\% of its life burning hydrogen and most of the
rest burning helium \citep[and references therein]{woosley02}; the
carbon, neon, oxygen, and silicon burning lasts about 1000 yr
altogether.  The star core temperature, matter density, the radius,
and the mass, all change during the evolution cycle. The mass-loss by
a star during its lifetime implies that stars with initial mass
$\ge35M_\sun$ and the solar metallicity to end their life as
hydrogen-free objects of roughly $5M_\sun$; early massive stars with
essentially no metallicity will evolve differently and may retain
a massive Hydrogen envelope even at later burning stages.

\begin{figure}[t]
\centerline{
\includegraphics[width=0.48\textwidth]{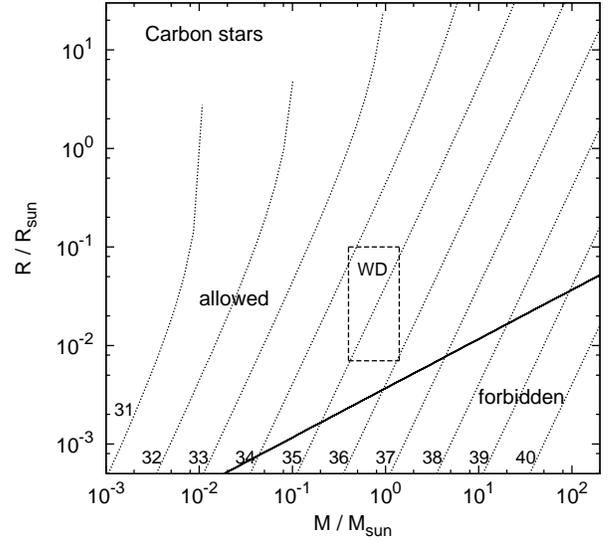}}
\caption{Mass--radius--capture rate diagram for Carbon stars.
The rectangular area outlined by the dashed line and labeled 
``WD'' shows parameter range typical for white dwarfs.
Lines are coded as in Figure~\ref{mapH}.
\label{mapC}}
\vspace{-1em}
\end{figure}

Figures~\ref{mapH}--\ref{mapC} show the mass--radius--capture rate
diagrams for $m_\chi=100$ GeV, $\sigv=3\times10^{-26}$ cm$^3$
s$^{-1}$, and WIMP velocity dispersion $\bar{v}=270$ km s$^{-1}$.  The
value $\rho_\chi=10^{10}$ GeV cm$^{-3}$ corresponds to the maximal
energy density allowed by the age of the supermassive black hole
$\sim$10 Gyr, and our selected values of $\sigv$ and $m_\chi$
\citep{gs99,bm05}.  The spin-dependent and spin-independent
WIMP-nucleon scattering cross sections are taken as
$\sigma_s=10^{-38}$ cm$^2$ and $\sigma_0=10^{-43}$ cm$^2$,
correspondingly. The series of dotted lines are calculated using
eq.~(\ref{C2}), and the numbers on the plots show the logarithm of
WIMP capture rate $\log_{10} C$.  The strong dashed line labeled
``MS'' in Figure~\ref{mapH} shows characteristics of the main sequence
stars \citep{encyclopedia}, and the diamond symbol indicates sun-like
star with $M=M_\odot$, $R=R_\odot$.  The open circles on the main
sequence show the stars of particular spectral type and the numbers
show their luminosity $\log_{10} (L_*/L_\odot)$.  The rectangular area
outlined by the dashed line and labeled ``WD'' in Figure~\ref{mapC}
shows the parameter range typical for white dwarfs.  The solid line
(eq.~[\ref{MR}]) divides the diagram into ``allowed'' and
``forbidden'' parts.  For any particular $M_*$ and $R_*$, the WIMP
capture rate can be found from the grid provided by the dotted lines.

Table \ref{table1} shows the WIMP capture rate, WIMP accumulation, and
other parameters for different star masses ($1-75 M_\odot$) and at
different evolution stages.  Spin-dependent WIMP accumulation is
calculated for the Hydrogen burning stage and spin-independent WIMP
accumulation is calculated for the Helium and Carbon burning stages.
The star masses, radii, core temperatures and densities are taken from
\citet{woosley02} and correspond to the particular burning stage. 
The $\tau_*$ column shows the duration of the corresponding burning cycle.
The uppermost row in the Carbon burning stage table shows the parameters
typical for a white dwarf where the core temperature is taken {\it ad
hoc}. This does not affect the WIMP capture rate, however. 
Smaller annihilation
cross section $\sigv<3\times10^{-26}$ cm$^3$ s$^{-1}$ will
lead to a larger number of captured WIMPs, although this would not
change the burning rate. Early massive low-metallicity stars can
potentially accumulate the most numbers of WIMPs due to the presence
of a large Hydrogen envelope.  Such stars could capture WIMPs at
greater rates due to the potentially larger spin-dependent
WIMP-nucleon scattering cross-section.
The WIMP number density may further increase during the sudden 
collapse of a massive star \citep{ullio01}.

In steady state the annihilation rate is equal to the capture rate. 
The energy release due to the WIMP annihilation in the star core is
\begin{equation}
L_\chi\sim 1.6\times10^{34} C_{35}
\left[\frac{m_\chi}{100\ \rm GeV}\right]\ {\rm erg\ s}^{-1},
\end{equation}
where $C_{35}$ is the capture rate in units $10^{35}$ s$^{-1}$. 

\section{Discussion and conclusion}

Where does the energy released during the WIMP annihilation go?  Table
\ref{table1} shows that the typical radius of the thermal distribution
of WIMPs in the star core $r_{th}\ll R_*$, therefore, the products of
their annihilation can not propagate to the star surface and are
eventually converted into the thermal energy and neutrino emission.
This may be not true for the white dwarfs whose radii are comparable
to $r_{th}$, and where the effect of gravitational ``focusing'' is 
important (eqs.~[\ref{C1}], [\ref{C2}], second term in square brackets).
While most of the energy is converted into heat,
the surface of white dwarfs may emit products of WIMP annihilation,
particles ($e^\pm$, $p$, $\bar p$, $\nu$, $\tilde{\nu}$) and \grays,
and this might be visible with a \gray\ telescope.

The WIMP capture rate generally increases with the star mass.
However, the relative importance of WIMP annihilation diminishes in
massive stars since their own luminosity increases as $\propto
M_*^{3.5}$. Therefore, the interiors of massive stars will
remain largely unaffected by WIMPs.

The WIMP burning affects mostly stars with $M\la M_\odot$ with low to
moderate luminosities and a large rate of WIMP capture.  The typical
luminosity of a sun-like star burning Hydrogen is
$L_\odot=3.8\times10^{33}$ erg s$^{-1}$, while WIMP annihilation in
its core can provide up to $L_\chi\sim4.6\times10^{34}$ erg s$^{-1}$,
i.e.\ 10 times more. Such an additional energy source may exceed
the star's own resources from nuclear burning.  The large
concentration of WIMPs may also lead to more effective energy
transport in the stellar core \citep{bouquet89} thus decreasing the
core temperature and limiting the replenishment of the burning core
with fresh nuclear fuel.  Combined, these two effects could change the
star's evolution scenario, life expectancy, and appearance.

A white dwarf, a star without its own energy supply consisting of
Carbon and Oxygen, may emit $L_\chi\sim2\times10^{33}-2\times10^{35}$
erg s$^{-1}$, i.e.\ 0.4--50 times luminosity of the sun, burning WIMPs
only and this energy source will last forever!  To reach such a
luminosity, the surface temperature of the white dwarf should be in
the range of $T\sim(80-150)\times10^3$ K assuming $R=0.01R_\odot$.  In
comparison, the luminosity of a regular white dwarf does not exceed
$\sim$$0.01L_\odot$. The interiors of white dwarfs are almost
isothermal, and the energy transport is dominated by degenerate
electrons \citep[see][for a recent review]{hansen04}; therefore, the
large number of captured WIMPs and their annihilation in the core
would not change the structure of white dwarfs. Interestingly,
since $L_\chi\propto\rho_\chi$, a population of such white dwarfs
located at different distances from the central black hole will
exhibit a luminosity correlated with the radial WIMP density
distribution.  A signature of a ``dark matter burner'' would be a
very hot star with a mass and radius characteristic for a white dwarf,
but with a luminosity exceeding the typical luminosity for a white
dwarf by orders of magnitude $\la$$50L_\odot$.  Note that an
independent determination of the $M_*/R_*$ ratio is possible using the
gravitational redshift which has to be equivalent to $v_R\sim50$ km
s$^{-1}$ \citep{gt67}.

Furthermore, a low-mass star with a highly eccentric orbit around the
central black hole may exhibit variations in brightness correlated
with the orbital phase.  The brightness should increase as the star
approaches the periastron and some time after that. To have this
working, the orbital period should essentially exceed
$\tau_{eq}$.  Carbon burning stars have $\tau_{eq}\sim10$ yr. It is
even shorter in case of a white dwarf $\tau_{eq}\sim0.5$ yr. 
If a white dwarf appears in a high-density WIMP region, the WIMP
density in its material quickly reaches equilibrium. On the other
hand, the small radii of white dwarfs $R_*\sim 10r_{th}$ imply that the
captured WIMP density close to the surface may be large enough thus
the surrounding WIMP density change will result in variations in their
brightness.  This makes white dwarfs ideal objects to test the WIMP
density in the environment in which they are orbiting.

Advances in near-IR instrumentation have made possible the
observation of stars in the inner parsec of the Galaxy
\citep{genzel00,ghez03,ghez05}.  The observed absorption line widths
imply high temperatures and lead to a ``paradox of youth:'' apparently
young stars in the region whose current conditions seem to be
inhospitable to star formation.  One of the possibilities is that they
are old stars masquerading as youths. If an independent determination
of their mass reveals that some of them are low-mass stars, then they
would become candidates for the dark matter burners.

The model predicts the existence of unusual stars and consequently
unusual supernovae in the central parts of the galaxies. These are the
stars burning WIMPs at a high rate whose luminosities essentially
exceed the typical luminosity of a star of given mass.
If found, such stars would be interesting probes of particle dark
matter near supermassive black holes. Their luminosity, or rather
their excess luminosity, attributed to the WIMP burning can be used to
derive the WIMP matter density at their location. On the other hand, a
lack of such unusual stars may provide constraints on WIMP density,
WIMP-nucleus scattering and pair annihilation cross-sections.

\acknowledgments

I.\ V.\ M.\ acknowledges partial support from NASA Astronomy and
Physics Research and Analysis Program (APRA) grant.  L.~Wai would like
to thank S.~Nagataki for interesting discussions on massive stars.
A part of this work was done at Stanford Linear Accelerator Center, Stanford
University, and supported by Department of Energy contract
DE-AC03-768SF00515.

\end{document}